\documentclass[aps,prb,twocolumn,showpacs,superscriptaddress]{revtex4}
\usepackage{graphicx}

\begin{document}

\title{Electronic specific heat in BaFe$_{2-x}$Ni$_x$As$_2$}

\author{Dongliang Gong}
\affiliation{Beijing National Laboratory for Condensed Matter Physics, Institute of Physics, Chinese Academy of Sciences, Beijing 100190, China}
\author{Tao Xie}
\affiliation{Beijing National Laboratory for Condensed Matter Physics, Institute of Physics, Chinese Academy of Sciences, Beijing 100190, China}
\author{Xingye Lu}
\affiliation{Beijing National Laboratory for Condensed Matter Physics, Institute of Physics, Chinese Academy of Sciences, Beijing 100190, China}
\author{Cong Ren}
\affiliation{Beijing National Laboratory for Condensed Matter Physics, Institute of Physics, Chinese Academy of Sciences, Beijing 100190, China}
\author{Lei Shan}
\affiliation{Beijing National Laboratory for Condensed Matter Physics, Institute of Physics, Chinese Academy of Sciences, Beijing 100190, China}
\affiliation{Collaborative Innovation Center of Quantum Matter, Beijing, China}
\author{Rui Zhang}
\affiliation{Department of Physics and Astronomy, Rice University, Houston, Texas 77005-1827, USA}
\author{Pengcheng Dai}
\affiliation{Department of Physics and Astronomy, Rice University, Houston, Texas 77005-1827, USA}
\author{Yi-feng Yang}
\affiliation{Beijing National Laboratory for Condensed Matter Physics, Institute of Physics, Chinese Academy of Sciences, Beijing 100190, China}
\affiliation{Collaborative Innovation Center of Quantum Matter, Beijing, China}
\author{Huiqian Luo}
\email{hqluo@iphy.ac.cn}
\affiliation{Beijing National Laboratory for Condensed Matter Physics, Institute of Physics, Chinese Academy of Sciences, Beijing 100190, China}
\author{Shiliang Li}
\email{slli@iphy.ac.cn}
\affiliation{Beijing National Laboratory for Condensed Matter Physics, Institute of Physics, Chinese Academy of Sciences, Beijing 100190, China}
\affiliation{Collaborative Innovation Center of Quantum Matter, Beijing, China}

\begin{abstract}

We have systematically studied the low-temperature specific heat of the BaFe$_{2-x}$Ni$_x$As$_2$ single crystals covering the whole superconducting dome. Using the nonsuperconducting heavily overdoped x = 0.3 sample as a reference for the phonon contribution to the specific heat, we find that the normal-state electronic specific heats in the superconducting samples may have a nonlinear temperature dependence, which challenges previous results in the electron-doped Ba-122 iron-based superconductors. A model based on the presence of ferromagnetic spin fluctuations may explain the data between x = 0.1 and x = 0.15, suggesting the important role of Fermi-surface topology in understanding the normal-state electronic states. 
\end{abstract}


\pacs{74.25.Bt,74.70.Xa}

\maketitle

\section{introduction}

The normal-state electronic states of the iron-based superconductors have been heavily studied by the specific heat technique \cite{StewartGR11}. The most important parameter is $\gamma_n = C/T$, which is proportional to the density of states at the Fermi energy and the effective electron mass. The measurements in the iron-based superconductors have given a variety of values of $\gamma_n$ for different materials \cite{MuG09,TanG13,GofrykK10a,GofrykK10b,HardyF10a,HardyF10b,PopovichP10,GofrykK11,PramanikAK11,HuJ11,ZengB12,NojiT12,JohnstonS14,StoreyJG13,AbdelM15}. A detailed study on the Ba(Fe$_{1-x}$Co$_x$)$_2$As$_2$ shows that $\gamma_n$ becomes maximum at optimal doping \cite{GofrykK10a,HardyF10b}, suggesting the influence of the antiferromagnetic (AF) order that disappears at the same doping \cite{DaiP15}. In addition to its relation to the normal-state electronic states, $\gamma_n$ is also associated with superconductivity though the ratio of $\Delta C/\gamma_nT_c$ that is traditionally used as an estimate of the coupling strength for a superconductor, where $\Delta C$ is the specific heat jump at the superconducting transition. The value of $\Delta C/\gamma_nT_c$ again peaks at the optimal doping \cite{HardyF10b}, suggesting strongest coupling of superconductivity around the optimal doping. 

Despite the abundant results shown above, the effect of spin fluctuations on the specific heat has been little studied. In a quasi-two-dimensional system close to an AF quantum critical point (QCP), $C/T$ may diverge logarithmically with decreasing temperature \cite{PetrovicC01,StewartGR01}. The iron-based superconductors undoubtedly exhibit strong AF fluctuations \cite{DaiP15}. Particularly, in the electron-doped BaFe$_{2-x}T_x$As$_2$ ($T$ = Ni or Co) system, it has been demonstrated that the long-range AF order is totally suppressed near optimal doping level, whereas the AF spin fluctuations survive in much higher doped samples, as shown in inelastic neutron scattering and nuclear magnetic resonance (NMR) measurements \cite{LuoH13,NingFL10,ZhouR13}. While the presence of an AF QCP in BaFe$_{2-x}T_x$As$_2$ is still under debate \cite{ZhouR13,LuoH12,LuX13}, the normal-state transport properties can be affected by the AF spin fluctuations around the optimal doping \cite{MayAF13,ArsenijevicS13,RullierAlbenqueF13}. Recently, ferromagnetic (FM) spin fluctuations have also been observed in the iron pnictides by the NMR technique \cite{WieckiP15}. These results clearly indicate that the effect of spin fluctuations should not be ignored in dealing with the specific-heat data. 

For the specific heat in the iron-based superconductors, phonon contribution is often dominated above $T_c$. To reveal the electronic specific heat, one may try to fit the data above $T_c$ with certain functions \cite{MuG09,TanG13} or find a reference sample without superconductivity by assuming that the specific heat of phonons change little with doping \cite{GofrykK10a,GofrykK10b}. In the latter case, the specific heat of the phonons of the reference sample may be simply adjusted as aC$_{phonon}(bT)$ with a and b as the tuning parameters (a-b method) to account for the change of the phonon spectra with doping \cite{HardyF10a,HardyF10b,PopovichP10,PramanikAK11,HuJ11,ZengB12,NojiT12,JohnstonS14,AbdelM15}. In most cases, a linear temperature dependence of the electronic specific heat below $T_c$ is assumed to account for the entropy conservation, {\it i.e.}, $\gamma_n$ is temperature-independent. An artificial linear or quadratic temperature dependence of $\gamma_n$ may also be chosen \cite{GofrykK10a,StoreyJG13}. Apparently, the non-linear temperature-dependent contribution to the specific heat from spin fluctuations is neglected.

In this paper, we systematically study the low-temperature electronic specific heat of the electron-doped BaFe$_{2-x}$Ni$_x$As$_2$ to address the role of spin fluctuations. Similar to other iron-based superconductors, the BaFe$_{2-x}$Ni$_x$As$_2$ system shows a superconducting dome with the total suppression of static AF order near the optimal doping level x $\sim$ 0.1 \cite{LuoH12,LuX13}. We find that using the non-superconducting heavily overdoped x = 0.3 sample as a reference for the phonon part of the specific heat without any artificial adjustment is sufficient to obtain the electronic specific heat in this system. For the samples with x between 0.1 and 0.15, $\gamma_n$ increases with decreasing temperature, which may be explained by the spin-fluctuation theory. Our results initiate that extra caution is necessary in studying the specific heat near the AF instability.

\section{experiments}

High-quality single crystals of BaFe$_{2-x}$Ni$_x$As$_2$ and BaFe$_{2-x-y}$Ni$_x$Cr$_y$As$_2$ were grown by the self-flux method as reported previously \cite{ChenY11,ZhangR14}. The superconductivity can be easily suppressed by a few percent of Cr doping \cite{ZhangR14}. In the following text, the pure Ni doped sample is denoted by only the value of x. To further simplify the description, the exact value of y will only be labeled in the figure captions for the non-superconducting Cr-doped samples. The actual and nominal doping levels of both Ni and Cr have linear relationships with the ratios of about 0.8 and 0.7, respectively. We will use the nominal values to be consistent with our previous reports \cite{LuoH12,LuX13,ZhangR15}. The specific heat was measured by the Physical Properties Measurement System (PPMS) from Quantum Design with or without He-3 option. The magnetic susceptibility was measured by the SQUID. 

\section{results}

\subsection{Validity of the a-b method}

Fig. 1(a) shows the whole specific heats of the BaFe$_{2-x}$Ni$_x$As$_2$ samples up to 30 K. Fig. 2(b) further shows the low-temperature data with $T^2$ as x axis. All the low-temperature data can be well fitted by a cubic equation as $C_{tot}/T = \gamma_0+BT+CT^2$ except for the x = 0.3 sample, where $C_{tot}$, $\gamma_0T$ correspond to the total and residual electronic specific heats, respectively. The slight deviation at very low temperature for the x = 0.3 sample may come from Schottky anomaly, which is absent in other samples. Fig. 1(c) and 1(d) show the doping dependence of coefficients B and C. The contribution of B$T^2$ has been attributed to line nodes or deep minima in the energy gap \cite{GofrykK11,MuG11,ZengB12}. The values of C around optimal doping are much larger than those in the nonsuperconducting underdoped and overdoped samples, suggesting that part of it should come from the electronic contribution such as point nodes in the superconducting gaps or some kind of bosonic mode \cite{GofrykK11}. In the x = 0.18 sample, the value of C is much smaller than that of other samples, most likely due to its low $T_c$, which makes the current upper bound of fitting temperature too high. Fitting the data with smaller temperature range results in larger value of C. Our results at low temperature are consistent with previous results in the electron-doped BaFe$_2$As$_2$ materials \cite{GofrykK11,MuG11,ZengB12}, suggesting the good quality of our samples. In this paper, we will focus on the study of the normal-state electronic specific heat.

Since the x = 0.3 sample shows neither superconductivity nor static AF order, it may be used as a reference to remove the phonon contribution. The presence of a large (~50 meV) spin gap in the x=0.3 sample \cite{WangM13} means that one can safely ignore the spin-fluctuation contribution to the specific heat in this material. In the following analysis, the phonon specific heat of the x = 0.3 sample is calculated as CT$^3$ where C is the fitted parameter as shown in Fig. 1(d). The phonon specific heat for the whole temperature range can be thus derived by subtracting $\gamma_n T$ from the total specific heat. Using $C_e^x = C_{tot}^x-aC_{phonon}^{0.3}(bT)$ where x and 0.3 represent doping levels, we can get the electronic specific heats $C_e^x$ of all the superconducting samples meeting the requirement of the entropy conservation by adjusting parameters a and b, as shown in Fig. 1(c). No adjustment is needed if both a and b are equal to one.

\begin{figure}[tbp]
\includegraphics[scale=.5]{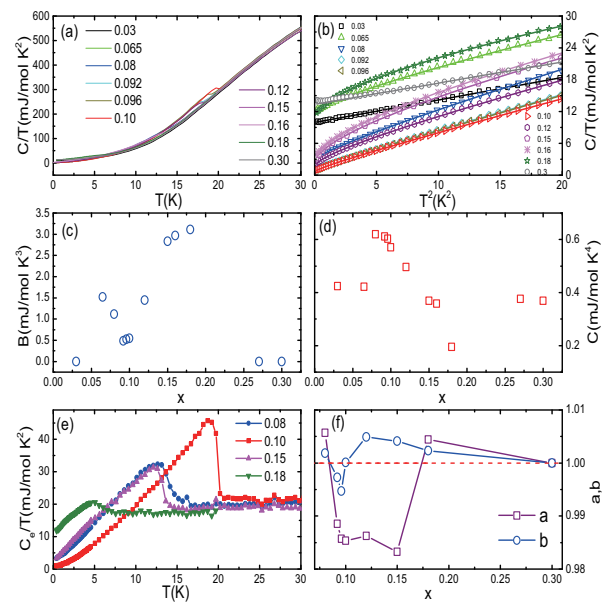}
\caption{ (a) Temperature dependence of the total specific heat of BaFe$_{2-x}$Ni$_x$As$_2$. (b) The $T^2$ dependence of low-temperature specific heat of BaFe$_{2-x}$Ni$_x$As$_2$. The solid lines are fitted by a cubic function as described in the text. (c) and (d) give the doping dependence of the fitting parameters B and C as discussed in the text, respectively. (e) The electronic specific heat of selected samples obtained by the a-b method. (f) The doping dependence of the parameters a and b in the a-b method. 
}
\label{fig1}
\end{figure}

The doping dependence of parameters a and b is shown in Fig. 1(d). At first glance, both a and b show only slight deviation from 1, suggesting the phonons do not change much with Ni doping as expected. Usually, it means that the data treatment is satisfactory.  However, we note that these two parameters do not change monotonically with doping. Specially, the value of b is supposed to only depend on the Debye temperature which should show monotonic dependence on Ni doping, whereas it changes abruptly around the optimal doping. Checking the details of the a-b method, the requirement of entropy conservation in the data treatment actually assumes that $\gamma_n$ is temperature independent. The nomonotonic Ni doping dependence of a and b suggests that such assumption may not be valid. Here we suggest that the a-b method assuming a temperature-independent $\gamma_n$ is indeed incorrect in this system based on the following arguments.

First, we find that different results will be achieved in some samples if we choose different temperature range to subtract the phonon contribution. Fig. 2(a) gives the subtracted $C/T$ of x = 0.15 sample obtained by considering different temperature ranges using the a-b method. The zero value of $C/T$ usually means that the adjustment by tuning a and b correctly captures the change of the specific heat from the phonon contribution. Large deviation from zero is found above 30 K if we only consider the data from 16 K to 30 K. On the other hand, while the fit looks better for the temperature range from 30 K to 60 K, the entropy does not conserve at $T_c$ as shown in the inset of Fig. 2(a). The inconsistency between the above two results suggests that the a-b method is not reliable. 

Second, the change of the phonon contribution to specific heat due to Ni dopings should be smaller than that suggested by Fig. 1(d). At low temperatures, the effect of doping on the specific heat of phonons may be directly associated with the atomic mass if the lattice structure is not changed. Fig. 2(b) gives the electronic specific heat of the x = 0.3, y = 0.3 sample using zero Cr doped sample as the reference without any adjustment. The actual atomic mass change is about 0.4\%, whereas that between the x = 0.3 and x = 0 samples is about 0.17\%. With much larger change of the atomic mass, the $C/T$ of the Cr-doped sample above 15 K suggests that no further adjustment is needed. The low temperature upturn below 15 K may be attributed to the Cr impurities \cite{ZhangR15}. Treated as a diluted spin system, we may fit the low-temperature data with the spin-fluctuation theory as follows \cite{TrainorRJ75,StewartGR84},

\begin{equation}
C = AT + BT^3 + DT^3lnT,
\label{eq1}
\end{equation}
\noindent as shown by the solid line in Fig. 2(b). The physical meanings of the parameters will be discussed later. We note that the $T_{SF}$ of the heavily Cr-doped sample in Fig. 2(b) is about 15 K, which is much smaller than the $T_{SF}$ shown in Fig. 4(a).

Third, the specific heat of slightly Cr-doped nonsuperconducting samples suggests that there is also low-temperature upturn of $C/T$ in the normal state of the superconducting samples. Fig. 2(c) gives the electronic specific heat of the x = 0.15 sample using the Cr-doped sample as a reference without any adjustment. The entropy is conserved as shown in the inset of Fig. 2(c), which suggests the reliability of the data treatment. While it indeed gives the correct electronic specific heat below $T_c$, the information on temperature dependence of the normal-state specific heat is missing since the zero value of $C_e/T$ above $T_c$ only suggests that the superconducting and nonsuperconducting samples have the same electronic specific heat assuming that the phonon specific heat changes negligibly with slight Cr doping. 

\begin{figure}[tbp]
\includegraphics[scale=.5]{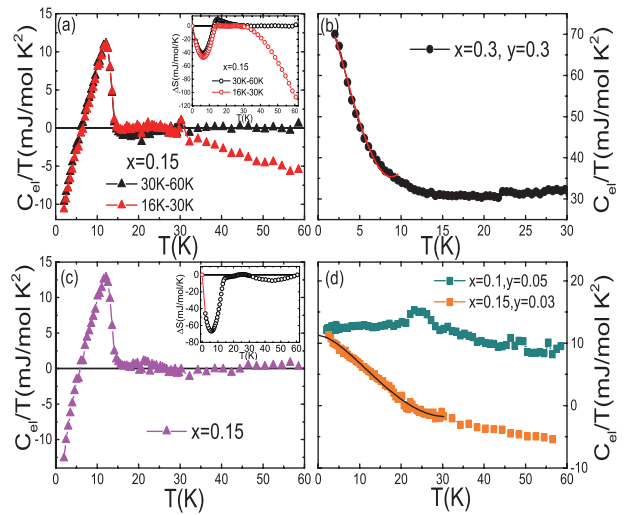}
\caption{ (a) The electronic specific heat of the x = 0.15 sample using the x = 0.3 sample as a reference adjusted by the a-b method, where the range of normal-state data is selected either from 30 K to 60 K (black triangles) or from 16 K to 30 K (red triangles). The entropy differences between the normal and superconducting states are shown in the inset. (b) The electronic specific heat of the x = 0.3, y = 0.3 sample using the x = 0.3 sample as the reference. The low temperature data are fitted by Eq. \ref{eq1} as shown by the solid line. (c) The electronic specific heat of the x = 0.15 sample using the nonsuperconducting x = 0.15, y = 0.03 sample as the reference. The inset shows that the entropy is conserved. (d) The electronic specific heat of the Cr-doped nonsuperconducting BaFe$_{1.85}$Ni$_{0.1}$Cr$_{0.05}$As$_2$ (squares) and BaFe$_{1.82}$Ni$_{0.15}$Cr$_{0.03}$As$_2$ (circles) using the x = 0.3 sample as the reference. The solid line is fitted by Eq. \ref{eq1}.
}
\label{fig2}
\end{figure}

Fig. 2(d) gives temperature dependence of the specific heat of the Cr-doped samples. For the Cr-doped x = 0.15 sample, $C/T$ increases quickly with decreasing temperature. Such increase of $C/T$ at low temperature cannot be compromised by the a-b method since the latter mainly affect the specific heat above 10 K. As shown in Fig. 2(c), the entropy conservation and the zero value of $C/T$ above $T_c$ suggests that $C/T$ of the x = 0.15 sample should also show upturn at low temperature if no superconductivity is present. In other words, the normal-state electronic specific heat of the x = 0.15 sample cannot be simply described by a linear temperature dependence. Similar to the heavily Cr-doped sample (Fig. 2(b)), we can also fit the low temperature data of the Cr-doped x = 0.15 sample with the spin-fluctuation theory. 

Unfortunately , using Cr-doped nonsuperconducting samples to obtain the normal-state electronic specific heat of the corresponding superconducting samples does not work for the samples below the optimal doping level. For example, the entropy is not conserved for the x = 0.1 sample using the x = 0.1 Cr-doped sample as the reference, suggesting that the electronic specific heats of the two samples are not the same. As shown in Fig. 2(d), $C/T$ of the x = 0.1 Cr-doped sample shows a kink around 25 K, which may be due to the enhancement of the AF order upon Cr doping \cite{ZhangR15}. 

\subsection{Electronic specific heat of BaFe$_{2-x}$Ni$_x$As$_2$}

\begin{figure}[tbp]
\includegraphics[scale=.4]{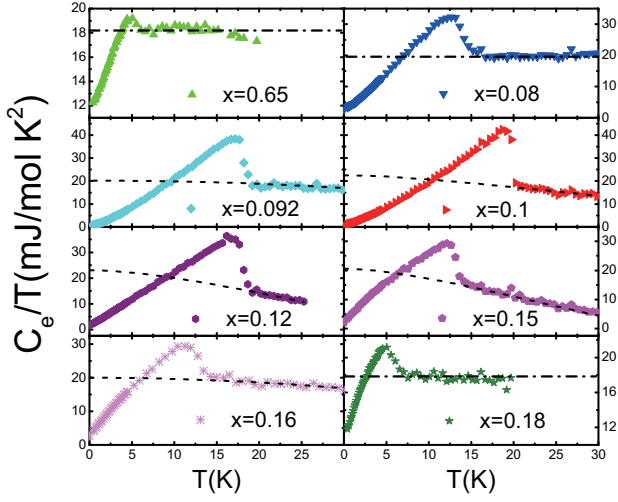}
\caption{ (a) - (h) The temperature dependence of the electronic specific heat of pure Ni-doped  samples with Ni doping levels as labeled in the figures using the x = 0.3 sample as reference without any adjustment. The dashed lines are fitted by Eq. \ref{eq1}. The dotted dashed lines are straight lines with constant value of $C/T$.
}
\label{fig3}
\end{figure}

The above results suggest that no adjustment is needed in deducting the phonon specific heat of x = 0.3 sample. Fig. 3 shows the electronic specific heat of superconducting samples with both a and b equal to one. A slightly decrease of $C/T$ with decreasing temperature above $T_c$ is found for the x = 0.08 sample, probably due to the presence of long-range AF order \cite{LuoH12,LuX13}. For the samples from x = 0.1 to 0.15, $C/T$ shows clear sign of upturn. While the normal-state electronic specific heat above $T_c$ can be obtained just by this simple subtraction, we may further obtain its value below $T_c$ without the presence of superconductivity using the spin-fluctuation theory with the constraint of entropy conservation, as shown by the dashed lines in Fig. 3. 

Fig. 4(a) and 4(b) shows the doping dependence of parameter $T_{SF}$ and $D$, where $T_{SF} = e^{-B/D}$ is the characteristic spin-fluctuation temperature \cite{TrainorRJ75,StewartGR84}. The weak upturns of $C/T$ in the x = 0.092, 0.096 and 0.16 samples cause unrealistic fitting parameters, such as negative $D$ or very large value of $T_{SF}$ ( about 10$^{24}$ K for the x = 0.16 sample). For the x = 0.1, 0.12, and 0.15 samples, while $T_{SF}$ shows weak doping dependence, $D$ increases linearly with doping. According to the spin-fluctuation theory, $D = \alpha\gamma/T_{SF}^2$ where $\alpha$ and $\gamma$ are associated with the Stoner-enhancement factor and the electronic specific-heat coefficient determined from the band-structure density of states. It will be interesting to see which factor dominates in determining the doping dependence of $D$. 

\begin{figure}[tbp]
\includegraphics[scale=.5]{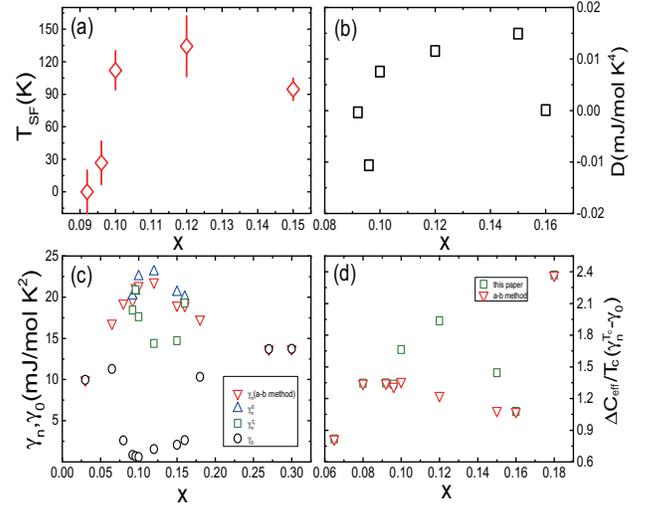}
\caption{ (a) Doping dependence of $T_{SF}$. (b) Doping dependence of $D$. (c) Doping dependence of $\gamma_n$ and $\gamma_0$. At zero K, two values of $\gamma_n$ obtained from the a-b method (red downward triangles) and our method (red triangles, $\gamma_n^0$) are given at each doping. $\gamma_n$ at $T_c$ is shown by the green squares ($\gamma_n^{T_c}$). (d) Doping dependence of $\Delta C_{eff}/T_c/(\gamma_n^{T_c}-\gamma_0)$ where $\gamma_n^{T_c}$ is obtained from either the a-b method (red downward triangles) or our method (green squares). 
}
\label{fig4}
\end{figure}

If there is indeed a non-linear temperature dependence of the normal-state electronic specific heat, one may ask how significant the effect is in determining the relevant physical properties. Fig. 4(c) gives the doping dependence of $\gamma_n$ and $\gamma_0$. The value of $\gamma_0$ gives the residual electronic specific heat coefficient at zero K, obtained by fitting the raw data as shown in Fig. 1(b). The very small values of $\gamma_0$ compared to $\gamma_n$ around the optimal doping level suggest the good quality of our samples. For the $\gamma_n$ at zero K, no distinct difference is found between the a-b method and our method. However, $\gamma_n$ at $T_c$ from x = 0.1 to 0.15 is significantly smaller than that of $\gamma_n$ at zero K. This results in large increases of $\Delta C_{eff}/T_c\gamma_n^{T_c}$ within the above doping range, where $\Delta C_{eff} = \Delta C\times\gamma_n/(\gamma_n-\gamma_0)$ assuming $\gamma_0$ comes from the non-superconducting part of the sample \cite{HardyF10a}. The maximum of the normalized superconducting jump shifts from the optimal doping level ( x $\approx$ 0.1 ) to slightly overdoped doping level ( x $\approx$ 0.12). The large deviation at x = 0.18 indicates that the large residual specific heat in heavily overdoped regime cannot be explained by phase separation, suggesting an inhomogeneously gapped superconducting state or pair breaking effect \cite{HardyF10a}. On the other hand, the quick increase of $\gamma_0$ with decreasing Ni for x $<$ 0.09 may come from the non-superconducting long-range antiferromagnetism. It should be noted that the determination of $\gamma_n^{T_c}$ is independent of any model including the spin-fluctuation theory discussed in the following section. 

\subsection{Magnetic properties of BaFe$_{2-x}$Ni$_x$As$_2$}

Fig. 5(a)-(f) show the magnetic susceptibility of BaFe$_{2-x}$Ni$_x$As$_2$ at various doping levels, where linear backgrounds at high fields have been subtracted. This is done by fitting the data between $\pm3000$ Oe to $\pm10000$ Oe as A$H$ $\pm\Delta M$, respectively, where A and $\Delta M$ are positive constants. Clearly ferromagnetic behavior at low fields is found for the samples from x = 0.1 to 0.15. Surprisingly, little change is seen between 25 K and 300 K. Fig. 5(g) gives the doping dependence of the saturate magnetic susceptibility at 25 K from the above results, which clearly suggests that the ferromagnetic moment becomes much stronger between x = 0.1 to 0.15. While these results are consistent with those of specific heat, it is the FM spin fluctuations that contribute to the electronic specific heat. Therefore, whether the observed ferromagnetism is directly associated with the enhancement of specific heat at low temperature is arguable. Fig. 5(h) further shows the temperature dependence of magnetic susceptibility at 7 Tesla for various samples, where an upturn at low temperature shows up in the x = 0.3 sample. However, such upturn seems to have negligible contribution to the specific heat as suggested by the nice fitting in Fig. 1(b). Similarly, the Cr impurities also introduce upturn in the magnetic susceptibility as shown in the inset of Fig. 5(h), but the electronic specific heats of the x = 0.15 sample with or without Cr doping are the same as discussed above.

\begin{figure}[tbp]
\includegraphics[scale=.45]{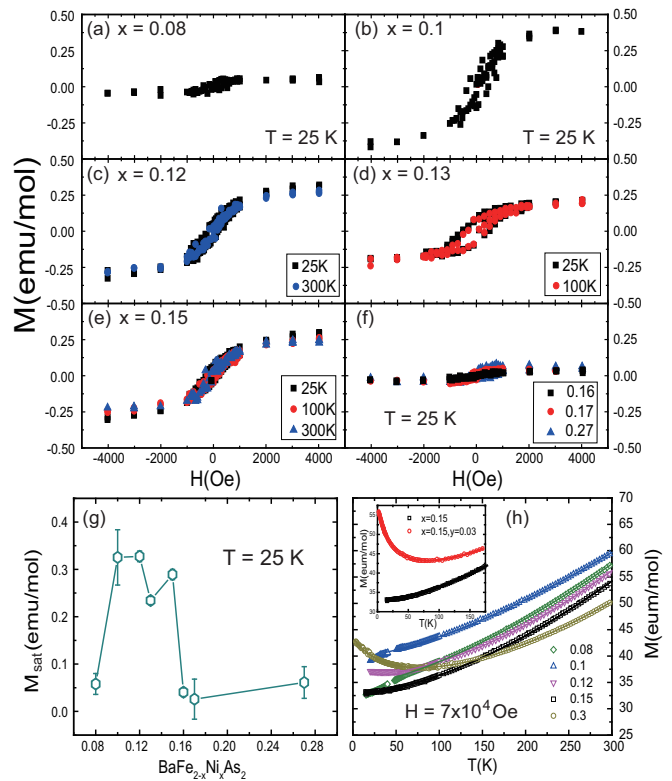}
\caption{ (a)-(f) Magnetic susceptibility of BaFe$_{2-x}$Ni$_x$As$_2$ samples with doping levels labeled in the figures after subtracting linear backgrounds measured at high fields as described in the text. (g) Doping dependence of saturated magnetic susceptibility at 25 K. (h) Temperature dependence of magnetic susceptibility for the pure Ni-doped samples at 7$\times$10$^4$ Oe. The inset shows the low temperature magnetic susceptibility for the x = 0.15 sample and the x = 0.15, y = 0.03 sample.
}
\label{fig5}
\end{figure}

\section{discussions}

From the above results of section III.A, we find that the a-b method cannot give the correct normal-state electronic specific heat in BaFe$_{2-x}$Ni$_x$As$_2$. In some of the studies \cite{HardyF10a,HardyF10b,PramanikAK11,JohnstonS14,AbdelM15}, only parameter a is introduced to account for the experimental uncertainties such as the error of sample mass. Since such process does not involve changing the temperature dependence of the phonon specific heat, it may not introduce some of the artificial results discussed above. However, we emphasis that the most important factor that determines applicability of the a-b or only-a method is whether the electronic specific heat has a linear temperature dependence, which is crucial in judging the entropy conservation. Moreover, if only a narrow range of normal-state specific heat is considered, the non-linear temperature dependence of the electronic specific heat can be easily neglected as shown in Fig. 2(a). 

It is rather surprising that we cannot observe the contribution to the specific heat from AF spin fluctuations. As shown by the neutron scattering and NMR experiments \cite{LuoH13,NingFL10,ZhouR13}, AF spin fluctuations should be dominant in this regime. Specially, the quantum critical spin fluctuations around the AF QCP can result in significant effect on specific heat \cite{PetrovicC01,StewartGR01}. However, as suggested by both NMR and neutron scattering measurements \cite{DioguardiAP13,LuX14}, the long-rang AF order in the electron-doped BaFe$_{2-x}T_x$As$_2$ system evolves into a cluster spin-glass state without the presence of AF QCP, which may explain the negligible contribution of AF spin fluctuations to the specific heat. On the other hand, the zero-$Q$ nature of FM spin fluctuations could have significant influence on specific heat. 

While the upturn of $C/T$ at low temperature for the samples with x from 0.1 to 0.15 can be reasonably explained by the spin-fluctuation theory, the origin of FM spin fluctuations needs to be further studied. Recently, the NMR measurement has found strong FM spin fluctuations in the Ba(Fe$_{1-x}$Co$_x$)$_2$As$_2$ system, which may be associated with our results. However, it seems to contradict with the observation that the low-temperature upturn of $C/T$ suddenly disappears above x = 0.16, as shown in Fig. 3. It should be noted that while $C/T$ of the x = 0.16 shows upturn in Fig. 3(g), the unrealistic fitting values of $T_{SF}$ and $D$ suggest that the effect of FM spin fluctuations should be weak and uncertainties such as masses of the samples may have to be considered. According to angle-resolved photo emission spectroscopy (ARPES) measurements \cite{LiuC11,IdetaS13}, such doping level corresponds to a topological change of the Fermi surfaces where the hole pocket at $\Gamma$ point disappears. The FM spin fluctuations contributed to the specific heat may thus come from the Stoner instability of the hole pocket at $\Gamma$ point, which should be suppressed by the AF order below x = 0.1 where the nesting of the hole and electron pockets sets in. Based on theoretical calculation \cite{SinghDJ08}, the mass enhancement around optimal doping is between 2 - 3. While the decrease of $\gamma_n$ at in the underdoped regime seems to be related to the Fermi surfaces reconstruction due to the presence of AF order \cite{HardyF10b}, the weakening of electronic correlation with Ni doping may cause the reduction of $\gamma_n$ in the overdoped regime \cite{YeZR14}. The spin-fluctuation contribution to the mass enhancement is 1 + $\frac{9}{2}$ln($S$/3) \cite{TrainorRJ75}, which gives value of the Stoner exchange-enhancement factor S between 3.7 to 4.7 around optimal doping level. It should be noted that the above analysis may be too simplified considering the multi-band and correlated nature of iron-based superconductors. 

Another possible origin of FM spin fluctuations is the presence of magnetic impurities \cite{VojtaT10}. Both Ni and Cr dopants may act as magnetic impurities as shown in Fig. 5(h) for the x = 0.3 and x = 0.15, y = 0.03 samples probably due to incomplete charge transfer \cite{TeagueML11}. The temperature dependence of magnetic susceptibility may be fitted by a Curie-Weiss-like plus a linear functions, the latter of which has already been observed in BaFe$_2$As$_2$ \cite{WangXF09}. The fitted mean-field transition temperatures for both samples are negative, suggesting an AF coupling. The low-temperature upturn of magnetic susceptibility has already been found in BaFeNiAs$_2$ \cite{SefatAS09a} and BaFe$_{2-x}$Cr$_x$As$_2$ system \cite{SefatAS09a}. Both Ni and Cr magnetic impurities have negligible effect on the specific heat as discussed previously. The doping evolution of the low-temperature upturn of magnetic susceptibility due to the Ni or Cr dopants does not shows a direct connection to that of the specific heat either. At current stage, we are unable to rule out the possibility that a very tiny amount of magnetic impurities may give rise to the enhancement of the specific heat at low temperature. The fact that the large ferromagnetic moment is only seen for the samples from x = 0.1 to 0.15 (Fig. 5(g)) also suggests that the magnetic impurities, if any, may not come from growth of the samples since they were grown with same process. 

In the end, we give a brief discussion on spin glass and its effect. It has been shown by neutron scattering experiments \cite{LuoH12,LuX14} that a spin-glass-like short-range incommensurate antiferromagnetism presents above x = 0.09 in BaFe$_{2-x}$Ni$_x$As$_2$. While the spin glass may give a FM-like hysteresis \cite{BinderK86}, it should show strong temperature dependence considering the $T_N$ measured by neutron scattering is just about 30-40 K around optimal doping, which contradicts our results in Fig. 5. Moreover, the spin glass state should die out quickly with increasing Ni doping above the optimal doping level, as shown in the similar Co-doped Ba-122 system \cite{DioguardiAP13}, but both the FM-like hysteresis and specific heat enhancement persist up to x = 0.15. Since the spin glass present at lower doping levels, it gives a natural explanation on the decrease of $\gamma_n^0$ below x = 0.12. The peak of $\Delta C_{eff}/T_c/(\gamma_n^{T_c}-\gamma_0)$ may thus be related to the phase fluctuations of the antiferromagnetic order, which shows glassy behavior around optimal doping. 

\section{conclusion}

In this paper, we have systematically studied the specific heat of the electron-doped iron-based superconductors BaFe$_{2-x}$Ni$_x$As$_2$. Our detailed analysis shows that the a-b method that is frequently used in the literatures cannot give the correct electronic specific heat for samples with x from about 0.1 to 0.15. The temperature dependence of the normal-state electronic specific heat within this doping range can be described by the spin-fluctuation theory, where the FM fluctuations may come from the Stoner instability of the hole pocket at $\Gamma$ point or a tiny amount of magnetic impurities. Our results suggest that the a-b method should be carefully used around the AF instability.

\acknowledgements

This work is supported by the "Strategic Priority Research Program (B)" of the Chinese Academy of Sciences (XDB07020300), the  Ministry of Science and Technology of China (973 project No. 2012CB821400, 2011CBA00110)  and the National Science Foundation of China (No. 11374011, 11374346, 91221303). The work at Rice University is supported by the
U.S. NSF-DMR-1362219, DMR-1436006, and in part by the Robert A. Welch Foundation Grant Nos. C-1839 (P.D.).

\bibliography{BFNASH12}

\end{document}